\documentclass[a4paper,12pt]{article}
\usepackage{epsfig,amsmath,amsfonts}
\newcommand{\eqsection}{\makeatletter
   \@addtoreset{equation}{section}
   \renewcommand{\theequation}{\arabic{section}.\arabic{equation}}
   \makeatother}

\begin{document}

\begin{center}
\LARGE{\bf {Exact Global Phantonical Solutions in the Emergent
Universe}} 
\end{center}

\begin{center}

Beesham A., \\

{\it {Department of Mathematical Sciences,  University of Zululand \\
Private Bag X1001, Kwa-Dlangezwa 3886, South Africa }\\
Email: abeesham@pan.uzulu.ac.za; Tel: +270762648833; Fax: +270312624852}

\bigskip
Chervon S.V.,\\%

{\it {Astrophysics and Cosmology Research Unit \\
School of Mathematics, Statistics and Computer Science, University of KwaZulu-Natal \\
Private Bag X54 001 \\Durban 4000, South Africa}} and

\sl{\it {Ulyanovsk State Pedagogical University named after I.N.
Ulyanov, Ulyanovsk 432700, Russia}}

\bigskip
Maharaj S.D., \\

{\it {Astrophysics and Cosmology Research Unit \\
School of Mathematics, Statistics and Computer Science, University of KwaZulu-Natal \\
Private Bag X54 001 \\Durban 4000, South Africa}}

\bigskip
Kubasov A.S., \\

\sl{\it {Ulyanovsk State Pedagogical University named after I.N.
Ulyanov, Ulyanovsk 432700, Russia}}

\end{center}

\small{We present new classes of exact  solutions for
an Emergent Universe supported by phantom and canonical scalar
fields in the framework of a two-component chiral cosmological
model. We outline in detail the method of deriving exact solutions, discuss the potential and kinetic interaction for
the model and calculate key cosmological parameters. We suggest
that this  this model be called a {\it phantonical Emergent Universe} because of the necessity
to have phantom and canonical chiral fields. The solutions obtained are valid for all time.}

\vskip 0.2em

\small{\it Keywords: Cosmology, Emergent Universe, Phantom Field,
Exact Solution}

\section{Introduction}
About a decade ago, Ellis \cite{ellmaa02,elmuts03} and co-workers
presented the idea of an emergent universe (EmU) which contains a
scalar field. Such models are interesting in that they start off
from an almost static state in the infinite past, hence avoiding
the initial singularity, and can accommodate late time
acceleration as well. There is no need for a quantum gravity era
and such a model can be made consistent with features known to us.
Since then (see refs. in \cite{bcmk12qm}), the EmU has been
studied within several contexts, including placing constraints on
the parameters of the model from observations. However, one
unpleasant feature of the original model is that the scalar field
diverges in the infinite past. In an earlier paper
\cite{bcmk12qm}, we analyzed the EmU scenario with two chiral
fields within the framework of a chiral cosmological model (CCM).
There it was found that one of the fields should be a phantom one
and the other a canonical one. Therefore we call this model a {\it
Phantonical EmU}.
Obtaining asymptotical solutions in \cite{bcmk12qm} gave us hope
of finding an exact one.
An example of such a solution, which solves the previously
mentioned problem, as well as being asymptotically reasonable, 
were presented in that work. In this article we present new
classes of exact solutions which include the previous one as a
special case.

The plan of the paper is as follows. In section 2, we present the
field equations, whereas in section 3 we show how to construct
exact solutions. In particular two classes of solutions are
presented, and the potentials and kinetic energies are discussed
in section 4. The key cosmological parameters of the models are
given in section 5, and the conclusion in section 6.

\section{The Phantonical EmU}

In the works \cite{bcmk12qm} and \cite{Beesham}   we presented
a general approach to the EmU from a CCM as the source of the
gravitational field. Here we will analyze the model with the
following basic equations. The metric of a homogeneous and
isotropic universe in the Friedman--Robertson--Walker (FRW) form
is
\begin{equation}\label{frw}
 d s^2=d t^2-a(t)^2\left(\frac{d r^2}{1-\epsilon r^2}+r^2d\theta^2+r^2\sin^2\theta
   d\varphi^2\right).
\end{equation}

We take the target space (chiral) metric in the form
\begin{equation}\label{tsm}
    ds_{\sigma}^2=h_{11}d\varphi^2 +h_{22}(\varphi,\psi)d\psi^2,~~~h_{11}=const.
\end{equation}

For the metrics (\ref{frw}) and (\ref{tsm}), the field equations
of the two component chiral cosmological model and Einstein's
equations can be represented in the form:
\begin{eqnarray}
\label{f-1}
h_{11}\ddot{\varphi}+3Hh_{11}\dot\varphi-\frac{1}{2}\frac{\partial
h_{22}}{\partial \varphi}\dot \psi^2+\frac{\partial
V}{\partial\varphi}=0,\\
\label{f-2} 3H(h_{22}\dot \psi)+\partial_t(h_{22}\dot
\psi)-\frac{1}{2}\frac{\partial h_{22}}{\partial \psi}\dot
\psi^2+\frac{\partial V}{\partial\psi}=0,\\
\label{e-1}
H^2=\frac{\kappa}{3}\left[\frac{1}{2}h_{11}\dot\varphi^2+
\frac{1}{2}h_{22}\dot\psi^2+V\right]-\frac{\epsilon}{a^2},\\
\label{e-2}
\dot H=-\kappa\left[\frac{1}{2}h_{11}\dot\varphi^2+
\frac{1}{2}h_{22}\dot\psi^2\right]+\frac{\epsilon}{a^2}.
\end{eqnarray}
This system of equations is a system of differential equations of
second order with three unknown variables: two chiral fields
$\varphi$ and $\psi$, and the potential $V$. In accordance with
the method of fine tuning of the potential \cite{chzhsh97plb}, the
law of evolution of the Universe  $a=a(t)$ is specified. The
metric of the target space is not fixed as is traditionally
accepted, giving us the freedom of adaptation to solving the
problem.
We will retain  $\epsilon =-1,0,+1$ throughout the paper to
keep possible an investigation for open universes if observations may
admit this possibility.

A nonlinear sigma model (NSM)
has already been considered as the source of the emergent universe 
\cite{Beesham}. Here we consider the 2-component NSM with the
potential of (self)interaction and we choose the scale factor in
the most general form \cite{Mukherjee06} as
\begin{equation}\label{a_emu}
a(t)=A(\beta+e^{\alpha t})^m,~~\alpha >0,~~\beta >0,~~m>0.
\end{equation}

The analysis of the general evolution of the EmU and physical
interpretation of the model's parameters was done in
\cite{bcmk12qm}. Let us only recall that the EmU started off from
the radius $a_i>M_P^{-1}$ \cite{ellmaa02} in the infinite past $t
\rightarrow -\infty $. Using this asymptote, we find that
$A=\frac{a_i}{\beta^m}$. Starting from the radius $a_i$, the scale
factor of the Universe then increases until the epoch $t=0$. Also
in \cite{bcmk12qm} we presented the general evolution of the total
kinetic energy and the potential.
Evolution of the equation of state was analyzed as well. We will
skip the solutions with asymptotic analysis and turn our attention
to the exact ones. Also we need to emphasize that the mappings
$\varphi (t), \psi (t)$ and $t(\varphi ), t(\psi)$ are single
valued and simple (not transcendental).

In accord with following description we will prescribe the
properties of the first field $\varphi $ to be a phantom one by
setting $h_{11}=-1$. But for the second field $\psi $ we suggest
a positive sign for the chiral metric component $h_{22}:h_{22}>0$. This
choice gives us the possibility of calling the model under consideration
(as it based on phant-o-m and can-o-nical fields) a {\it phantonical
EmU}.

\section{The method of exact solutions construction. Two classes of exact solutions}

In the article \cite{bcmk12qm} we found an example of an exact
solution when considering the asymptotic regime. Now we will
present a method of constructing exact solutions.

We are looking for solutions of the equations
(\ref{f-1})-(\ref{e-2}) with the given scale factor
(\ref{a_emu}). This gives us the opportunity to calculate the
potential with
\begin{equation}\label{v-tot}
  V(t)=\frac{3}{\kappa}\left(H^2+\frac{1}{3}\dot
    H+\frac{2}{3}\frac{\epsilon}{a^2}\right),
\end{equation}
and the kinetic energy as well

\begin{equation}\label{kinetic}
  \frac{1}{2}h_{11}\dot\varphi^2(t)+\frac{1}{2}h_{22}(t)\dot\psi^2(t)=
  \frac{1}{\kappa}\left[\frac{\epsilon}{a^2}-\dot
    H\right].
\end{equation}
The formulas (\ref{v-tot}) and (\ref{kinetic}) are nothing else but
consequences of equations (\ref{e-1}) and (\ref{e-2}) which can be
obtained by making a simple algebraic conversion of the Einstein
equations. This is an analogue of fine tuning of the potential
method applied to a single scalar field \cite{chzhsh97plb}.

Let us start with a simplification for the chiral metric
components
\begin{equation}\label{h-1}
h_{11}=-1,~~h_{22}=h_{22}(\varphi). 
\end{equation}
We choose the potential  in the following form
\begin{equation}\label{V-dcm}
V(\varphi, \psi)=V_1(\varphi) + F(\varphi)V_2(\psi).
\end{equation}
From this form we decompose the total potential (\ref{v-tot}) for
the two parts as
\begin{eqnarray}
\label{v-1-1}
V_1(\varphi(t))=\frac{3}{\kappa}\left(H^2+\frac{1}{3}\dot
H\right),\\
\label{v-2-1}
F(\varphi(t))V_2(\psi(t))=\frac{2}{\kappa}\frac{\epsilon}{a^2}.
\end{eqnarray}
The kinetic energy also will be decomposed into two parts:
\begin{eqnarray}
\label{k-1-1} h_{11}\dot\varphi^2=-\frac{2}{\kappa}\dot H ,\\
\label{k-2-1}
h_{22}\dot\psi^2=\frac{2}{\kappa}\frac{\epsilon}{a^2}.
\end{eqnarray}

Taking into account that the first field $\varphi $ is a phantom
one, that is $h_{11} =-1$, one can derive the solution for the
phantom field

\begin{equation}\label{vp-ex}
\tilde{\varphi}:=\frac{\varphi(t)-\varphi_i}{2\sqrt{\frac{2m}{\kappa}}}=
\arctan\left(\frac{e^{\frac{\alpha}{2}t}}{\sqrt{\beta}}\right).
\end{equation}
Here we omitted the second sign for the square root because of the
necessity to have single valued functions mentioned earlier.
With the kinetic energy decomposition (\ref{k-1-1})-(\ref{k-2-1})
we can split the field equation (\ref{f-1}) in the following way
\begin{eqnarray}
\label{f-1-1}
-\ddot{\varphi}-3H\dot\varphi+\frac{\partial
V_1(\varphi)}{\partial\varphi}=0, \\
\label{f-2-1} -\frac{1}{2}\frac{\partial h_{22}}{\partial
\varphi}\dot \psi^2+V_2(\psi)\frac{\partial F(\varphi)}{\partial
\varphi}=0.
\end{eqnarray}
Thus the term $\frac{\partial V_1(\varphi)}{\partial\varphi} $ can
be reconstructed from equation (\ref{f-1-1}). The partial
derivative $\frac{\partial V(\varphi)}{\partial\varphi}$ can be
calculated as the sum: $\frac{\partial
V(\varphi)}{\partial\varphi}=\frac{\partial
V_1(\varphi)}{\partial\varphi}+V_2(\psi)\frac{\partial
F(\varphi)}{\partial \varphi}$. The total derivative
$\frac{dV(\varphi,\psi)}{d t}$ is calculated from the relation
\begin{equation}\label{V_tot_t}
\frac{dV(\varphi,\psi)}{d t}= \frac{\partial
V_1(\varphi)}{\partial\varphi}\dot{\varphi}+V_2(\psi)\frac{\partial
F(\varphi)}{\partial \varphi}\dot{\varphi}+ \frac{\partial
V_2(\psi)}{\partial\psi}\dot{\psi}.
\end{equation}
The first term on the rhs, $\frac{\partial
V_1(\varphi)}{\partial\varphi}\dot{\varphi}=:\dot{V_1} $, should
satisfy the relation
\begin{equation}
\dot{V_1}(\varphi, \psi)=\ddot H+6H\dot H,
\end{equation}
obtained by differentiating (\ref{v-1-1}) with respect to $t$.

Let us turn our attention to the second part of the field equation
(\ref{f-1}), viz., equation (\ref{f-2-1}). By inserting
$\dot{\psi}^2$ from (\ref{k-2-1}) into (\ref{f-2-1}) and using
(\ref{v-2-1}) one can derive
\begin{equation}\label{h-F}
\sqrt{h_{22}(\varphi)}=F(\varphi ),~~ const=1.
\end{equation}
(Here we set the integration constant equal to unity.)
Using (\ref{v-2-1}) we can rewrite the equation for the second
chiral field $\psi $ in the following way
\begin{equation}\label{f-2-simpl}
3Hh_{22}\dot\psi+\partial_t(h_{22}\dot\psi)+F\frac{\partial
V_2}{\partial\psi}=0.
\end{equation}
Multiplying this equation by $\dot\psi,$ we transform the last
equation to the time dependence
\begin{equation}\label{f-2-simpl-t}
3Hh_{22}\dot\psi^2+\partial_t(h_{22}\dot\psi)\dot\psi+F\frac{\partial
V_2}{\partial t}=0.
\end{equation}
Using (\ref{v-2-1}), (\ref{h-F})  equation (\ref{f-2-simpl-t})
reduces to
\begin{equation}\label{psi-l}
 \frac{2}{\kappa}\frac{\epsilon}{a^2}
 \left(H+\frac{\dot{F}}{F}\right)+F^2\dot{\psi} \ddot{\psi}=0.
\end{equation}
It is easy to check that this equation satisfies relation
(\ref{k-2-1}).

Thus we obtained a class of exact solutions for the phantonical
EmU supported by two dark sector fields, which can be described by
the formulas (\ref{h-1})-(\ref{vp-ex}).

\subsection{The first class of exact solutions}

The class of exact solutions for the model described by equations
(\ref{f-1})-(\ref{e-2}) with the Hubble parameter corresponding to
the EmU (\ref{a_emu}) is represented
by the formulas: (\ref{h-1}), (\ref{vp-ex}).
In addition, the second field $\psi $ and the kinetic interaction
term $ h_{22}$ should satisfy equation (\ref{k-2-1}).
The total potential $V(\varphi,\psi )$ should satisfy the
decomposition (\ref{V-dcm}), where

\begin{eqnarray}
\label{v1-1}
&V_1(\varphi)= \frac{m\alpha^2}{4}\sin^2(2\tilde{\varphi}
)[3m\tan^2(\tilde{\varphi})+1],\\
\label{h-F-f}
&F(\varphi )=\sqrt{h_{22}(\varphi)}, \\
\label{v2-1}
&V_2(\psi (t))=\sqrt{h_{22}}\dot\psi^2.
\end{eqnarray}
To be more illustrative let us suppose that the second field $\psi
$ is proportional to the cosmic time. Then expressing the time
inverse $ \varphi$ from (\ref{vp-ex}), and using the decomposition
(\ref{v-1-1})-(\ref{v-2-1}), we obtain an example of an exact
solution

\begin{eqnarray}
&\psi=\sqrt{\frac{2\epsilon}{\kappa}}t +\psi_0, \\
&h_{22}(\varphi)=\frac{[\cos^2(\tilde\varphi)]^{2m}}{a_i^2},\\
&V_2(\psi)=\frac{2}{A}\left(\beta +\exp\left[\alpha
\sqrt{\frac{\kappa}{2\epsilon}}(\psi-\psi_0)\right]\right)^{-m},
\end{eqnarray}
together with (\ref{vp-ex}) and (\ref{v1-1}).

This is the example of the solution we obtained earlier in
\cite{bcmk12qm}. The extension of the solution may be constructed
if we know the kinetic interaction between the dark sector fields.
It was shown \cite{arrche12gc} that exponential interaction
$h_{22}\propto \exp (-K_1 \tilde{\varphi}), K_1=const $ is
admitted. But for this kinetic interaction it is difficult to carry out a
numerical integration without a knowledge of the model's parameters
included in (\ref{a_emu}).

We should mention here one special solution belonging
to the first class of exact solutions, viz., the phantom field
$\varphi $ as a single field can support a spatially flat EmU
without an additional canonical field $\psi $. This solution is
described by $\psi = const=0 $ which means a degeneration of the
chiral metric $h_{AB}$ to one dimension. The evolution of the phantom field
 is given by (\ref{vp-ex}), the potential by
(\ref{v1-1}) with $h_{22}=0$ in (\ref{h-F-f}) and consequently
with $V_2=0$ in (\ref{v2-1}).

\subsection{The second class of exact solutions}

Another class of exact solutions of the EmU can be obtained with
the following assumptions on the kinetic and potential energy
forms. Let us assume a dependence of the chiral metric
component $h_{22}$ on the second field $\psi $. This case may be
considered as somewhat artificial. Indeed we can introduce a new
field by the relation $d\tilde{\psi} = \sqrt{h_{22}}d\psi $ and
the metric component $h_{22}$ will be absorbed. But there are two
arguments for this presentation. Firstly, it may help us in
searching for exact solutions because of the integral
transformation for $\tilde\psi $ above. Secondly, we can include the
crossing of the phantom zone if needed.

Let us introduce the following representations for the kinetic and
potential energy

\begin{equation}
h_{22}=h_{22}(\psi),~~V(\varphi,\psi)=V_1(\varphi)+V_2(\psi).
\end{equation}
Then we set the connection with cosmological dynamics and
curvature for the potential parts:
\begin{eqnarray}
\label{V-sub}
V_1(\varphi)= \frac{1}{\kappa}\left(3 H^2+\dot{H} \right),\\
\label{V-sub-2}
V_2(\psi)=\frac{2}{\kappa a^2}.
\end{eqnarray}
The connections for the kinetic parts are:
\begin{eqnarray}
\label{K-sub}
h_{11}{\dot{\varphi}}^2= -\frac{2}{\kappa}\dot{H},\\
\label{K-sub-2}
h_{22}{\dot{\psi}}^2 =\frac{2}{\kappa a^2}.
\end{eqnarray}
By direct substitution one can check that
(\ref{V-sub})-(\ref{K-sub-2}) are sufficient for the validity of
the Einstein and chiral field equations (\ref{f-1})-(\ref{e-2}).

Thus we can formulate the method of generation of exact solutions
of the second class. For a given scale factor evolution (in our
case $a(t)=A(\beta+e^{\alpha t})^m $)  we can solve the ansatz for
the first field $\varphi$ in (\ref{K-sub}).
The solution is represented by (\ref{vp-ex}).
Replacing $t$ by $\varphi $ in (\ref{vp-ex}) we obtain the first
term of the potential from (\ref{V-sub}) in the form (\ref{v1-1}).
The second part of the potential $V_2$ can be reconstructed from
(\ref{V-sub-2}) if we know the dependence $\psi =\psi (t)$.
Besides, the field $\psi $ and kinetic interaction term $ h_{22}$
should satisfy equation (\ref{K-sub-2}).

To illustrate the method let us assume once again the linear
evolution in time of the second field $\psi $

\begin{equation}\label{psi-2}
\psi(t)=\sqrt{\frac{2\epsilon}{\kappa}}t+\psi_0.
\end{equation}
Then we can define the metric component $h_{22}$ in terms of
cosmic time $t$ from (\ref{K-sub-2}) and then transform it to the
dependence on $\psi $ with (\ref{psi-2}):

\begin{equation}\label{h22-2}
h_{22}(\psi)=A^{-2}\left(\beta+
\exp{\alpha\left(\sqrt{\frac{\kappa}{2\epsilon}}(\psi-\psi_0)\right)}\right)^{-2m}.
\end{equation}
Using once again (\ref{psi-2}) we obtain the second part of the
potential

\begin{equation}\label{v2-2}
V_2(\psi)=\frac{2\epsilon}{\kappa}\,{A^{-2}\left(\beta+
\exp{\alpha\left(\sqrt{\frac{\kappa}{2\epsilon}}(\psi-\psi_0)\right)}\right)^{-2m}}.
\end{equation}
Let us mention here that for the exponential or logarithmic
evolution of the second field $\psi $ it is not difficult to
calculate the kinetic interaction term $h_{22}$. The results of
calculations will extend the list of exact solutions.

To analyze the obtained classes of exact solutions we turn our
attention to the potential and kinetic energy as a function of two
fields. We will also calculate the cosmological parameters for the
phantonical EmU to constrain them from observations.

\section{The potentials and kinetic energy}

Let us consider the total potentials for the solutions. The total
potential for the first class of solutions, represented by
formulas (\ref{v1-1})-(\ref{v2-1}) with their substitution in
(\ref{V-dcm}), is
\begin{equation}\label{vtot-1}
V(\varphi,\psi)=\frac{1}{\kappa}\left[\frac{m\alpha^2}{4}\sin^2(2\tilde\varphi)
[3m\tan^2(\tilde\varphi)+1]+ \frac{2\epsilon}{A a_i}
\frac{\cos^{2m}(\tilde\varphi)}
{(\beta+\exp{\alpha\left(\sqrt{\frac{\kappa}{2\epsilon}}(\psi-\psi_0)\right)})^m}\right].
\end{equation}
This solution is displayed in Fig. 1. We can find two regimes
with a behavior for the  $\varphi $-fixed field while the potential
$V$ is close to a constant value $V_*$ when $|\psi | > 10$.
Nevertheless in this regime one can obtain $H \propto \tanh
(\sqrt{3V_*}t)$ and so $a \propto \left[\cosh (\sqrt{3V_*}t)
\right]^{1/3}$. This solution means that the EmU will decay to an
inflationary type evolution far from the inflationary time. So this
regime is not important for the EmU scenario.

An analogous situation holds for the second solution (Fig. 2). The formula for calculating the potential  is

\begin{equation}\label{vtot-2}
V(\varphi,\psi)=\frac{1}{\kappa}\left[\frac{m\alpha^2}{4}\sin^2(2\tilde\varphi
)[3m\tan^2(\tilde\varphi)+1]+
\frac{2\epsilon}{A^2\left(\beta+\exp{\alpha(\sqrt{\frac{\kappa}{2\epsilon}}(\psi-\psi_0))}\right)^{2m}}\right].
\end{equation}
The differences with Fig. 1 are due to the
$\cos^{2m}(\tilde\varphi)$ in the numerator of the second term in
(\ref{vtot-1}) and the value of the coefficient and the power
($2m$) in the denominator of the second term in (\ref{vtot-2}).

\begin{figure}[h!]
\center
\includegraphics[width=3.25 in]{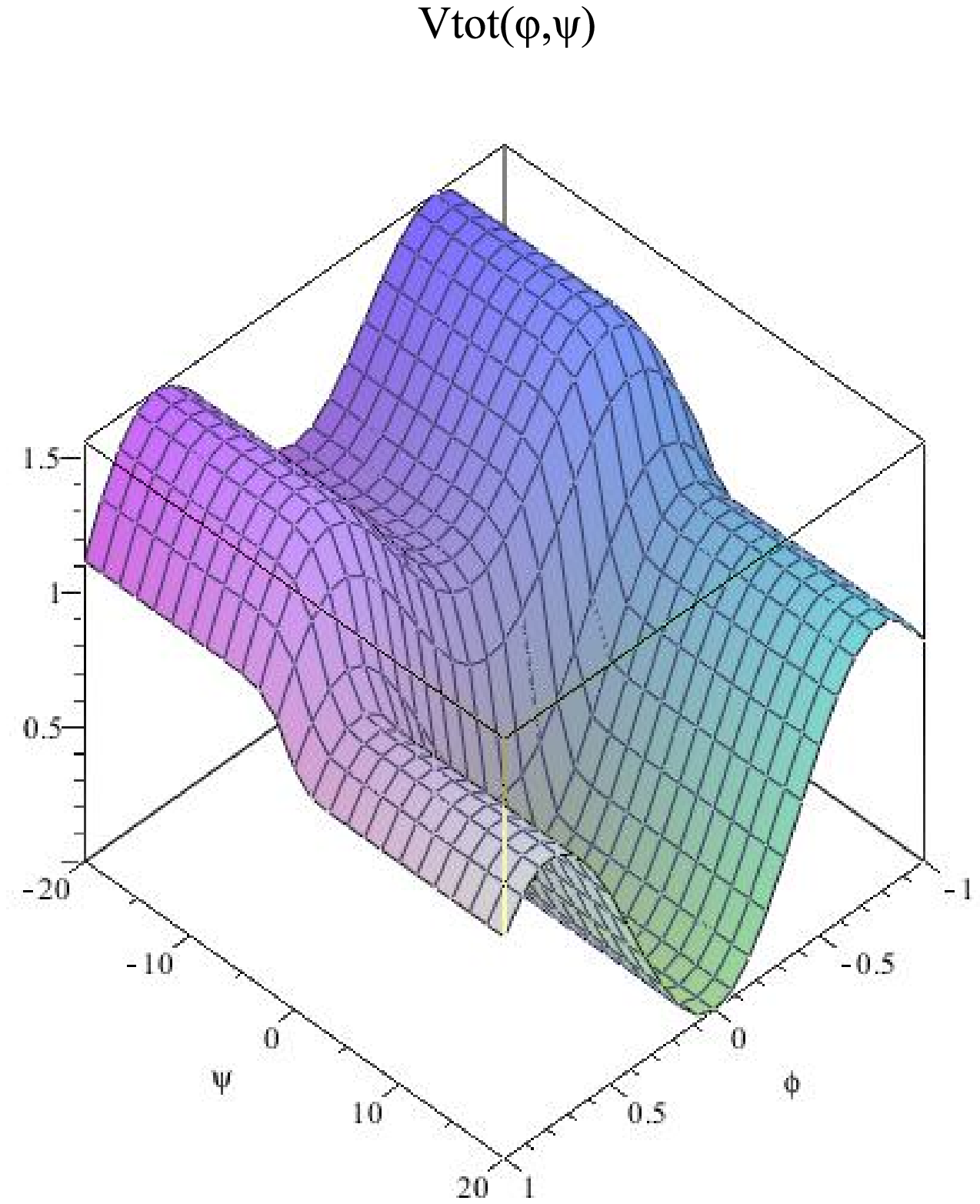}
\caption{The potential vs the field $\varphi, \psi $ for solution 1}
\label{ris:1}
\end{figure}

\begin{figure}[h!]
\center
\includegraphics[width=3.25 in]{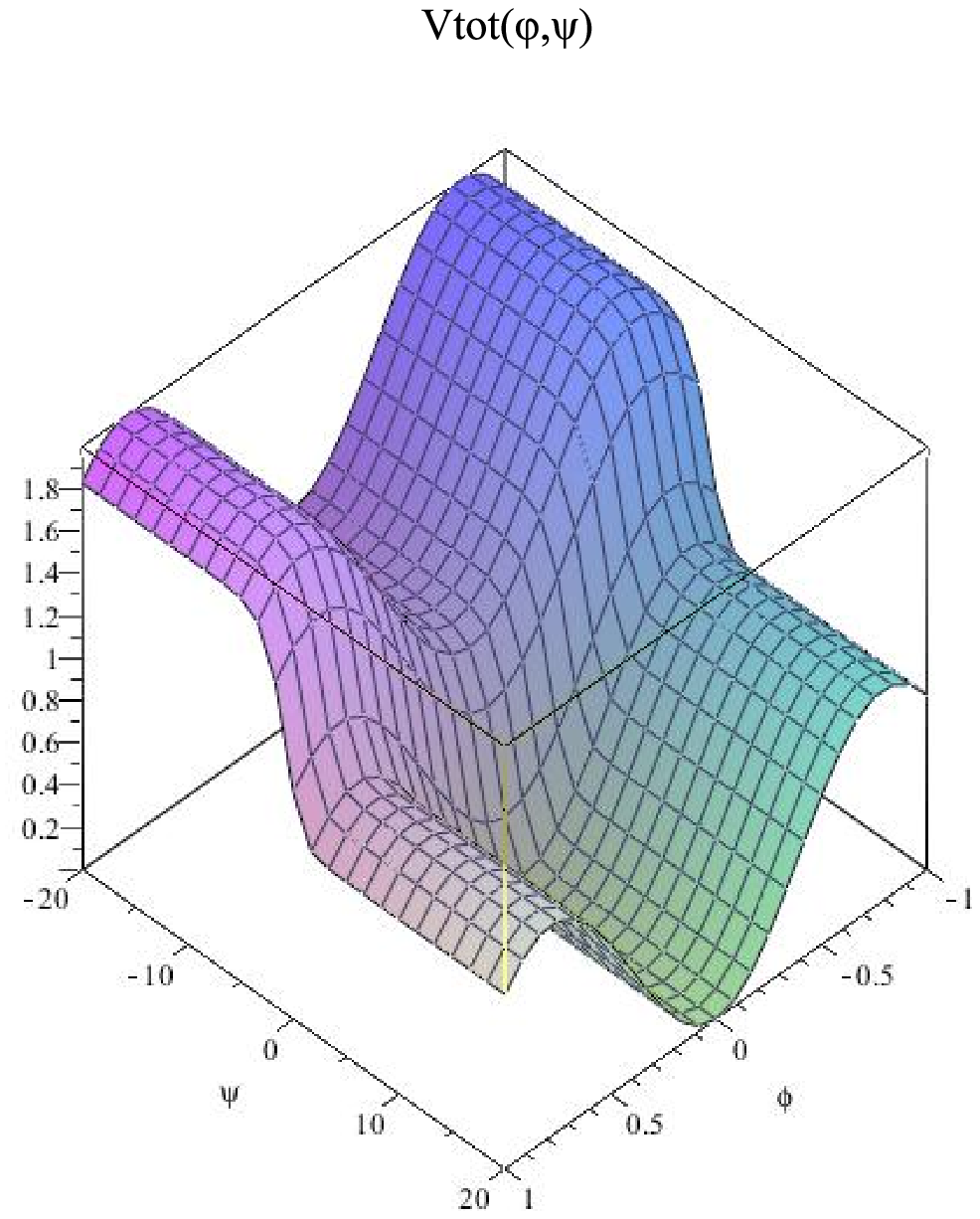}
\caption{The potential vs the fields $\varphi, \psi $ for solution 2}
\label{ris:2}
\end{figure}

The evolution of the kinetic energy  for the solutions is
presented in Fig. 3 and Fig. 4. The kinetic energy for the first
solution (Fig. 3) is

\begin{equation}\label{ktot-1}
K(\varphi)=\frac{1}{\kappa}\left[-\frac{m\alpha^2}{4}\sin^2(2\tilde\varphi)+
\epsilon\frac{(\cos(\tilde\varphi))^{4m}}{a_i^2}\right].
\end{equation}
We note the local maximum at an inflationary period ($ t \sim 0
$). The kinetic energy for the second solution (Fig. 4)
\begin{equation}\label{ktot-2}
K(\varphi,\psi)=\frac{1}{\kappa}\left[-\frac{m\alpha^2}{4}\sin^2(2\tilde\varphi)+
\frac{\epsilon}{A^2}\left(\beta+\exp{\alpha\left(\sqrt{\frac{\kappa}{2\epsilon}}
(\psi-\psi_0)\right)}\right)^{-2m}\right],
\end{equation}
has a local maximum in respect of the phantom field $\varphi $ during the
inflationary period.

\begin{figure}[h!]
\center
\includegraphics[width=3.25 in]{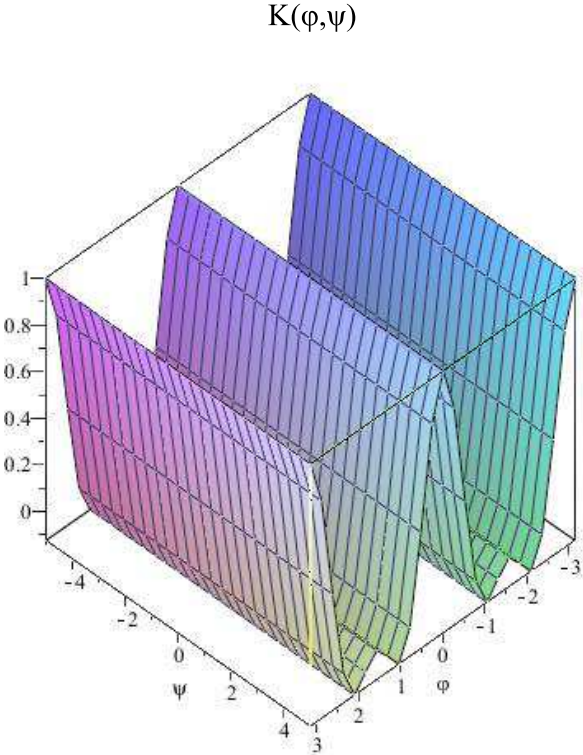}
\caption{The kinetic energy vs the field $\varphi $ for solution 1}
\label{ris:3}
\end{figure}

\begin{figure}[h!]
\center
\includegraphics[width=3.25 in]{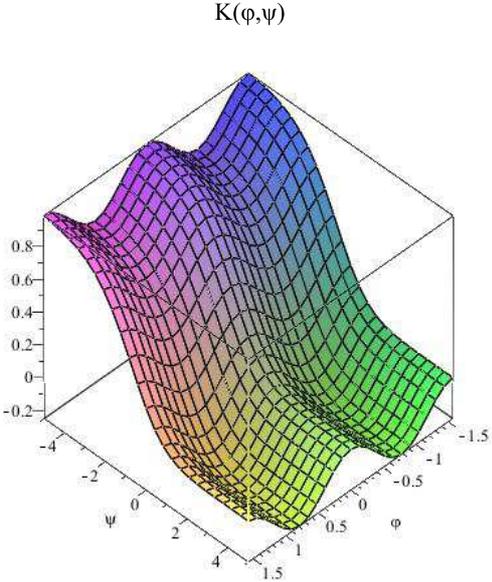}
\caption{The kinetic energy vs the fields $\varphi, \psi $ for
solution 2} \label{ris:4}
\end{figure}

\section{Key cosmological parameters of EmU}

We have obtained two classes of exact solutions and therefore it
is possible to apply the method of calculation of the cosmological parameters
 for this case \cite{chefom08}. We calculate the
parameters in accordance with the general approach at the time when the
perturbation with wave vector $\vec{k}$ is crossing the horizon:
$aH=k$. Let us denote by $t_*$ the time of horizon crossing. Then
the results for scalar type perturbations is presented by a power
spectrum of  the form
\begin{equation}
P_R(k)=\frac{m^3\alpha^2e^{3\alpha
t_*}}{8M_{p}^2\beta(\beta+e^{\alpha t_*})^2},
\end{equation}
and by spectral index

\begin{equation}
n_S(k)-1=\frac{3\beta+e^{\alpha t_*}}{me^{\alpha t_*}+\beta}.
\end{equation}
The power spectrum for tensor perturbations (gravitational waves) is

\begin{equation}
P_G(k)=\frac{m^2\alpha^2e^{2\alpha t_*}}{2M_{p}^2(\beta+e^{\alpha
t_*})^2}.
\end{equation}
The spectral index for tensor perturbations is

\begin{equation}
n_G=\frac{2\beta}{me^{\alpha t_*}+\beta}.
\end{equation}
Also we can calculate the tensor to scalar ratio (T/S ratio) as

\begin{equation}
\frac{T}{S}=4\frac{\beta}{me^{\alpha t_*}}.
\end{equation}
In the formulas above, $t_*$ is the time of horizon crossing.
These formulas can be used to make some restrictions for the EmU
parameters by comparing cosmological parameters with observational
data. But this procedure lies outside the framework of this
article and is a topic for future work.

\section{Discussion}

An EmU scenario is of great interest because of the possibility of
avoiding the initial singularity and a quantum
gravity era. Various aspects of the EmU scenario have been discussed
in the literature (see, e.g., \cite{bcmk12qm}). In the present
article, we have continued our  investigations connected with the
consideration of a NSM as the source of the EmU started in the
work \cite{Beesham}. There, a  two-component NSM was presented as
the source supporting the EmU development in time. The study was
performed for special regimes: early times ($t\rightarrow -\infty
$) and inflationary expansion.  There were found solutions
especially for these regimes.

As the next step of the investigation, we studied \cite{bcmk12qm}
the two-component CCM without any restriction
to special regimes; the entire evolution of the EmU was analyzed.
a few asymptotic solutions were found and some facts about the kinetic
interaction between dark sector fields were obtained via
an analysis of the chiral metric components $h_{22}$ (which is responsible for the
kinetic interaction in that model).
An example of an exact solution was found also which gave us hope
to describe a global evolution of the EmU from $t\rightarrow
-\infty $  to the infinite future and to calculate the
cosmological parameters for comparison with observational data.

In the present letter we report two new classes of exact solutions
which are valid for the entire evolution of the Universe. The
general features of the obtained solutions provide a strong
indication that  two types of scalar fields are needed: one field
should be of the phantom type while the other should be a
canonical one. Thus we arrive at a  generalization of the quintom
model \cite{chervon12qm} and with the arguments above, we suggest that
 this model be called a {\it phantonical EmU}. Therefore only such a
distribution between the fields will protect the EmU from decay
and give us the possibility to have  global evolution.

By analyzing the total potentials for both solutions (Fig. 1 and
Fig. 2), we find an analogy with the potential for hybrid inflation
\cite{linde94prd}. 
But if we will follow some regime with say $\psi = constant $,
the result will show a decay of the scale factor from the EmU one
(\ref{a_emu}).

Let us also pay attention to the presently observed  accelerated
expansion of the Universe. In the framework of the scale factor (\ref{a_emu}), we
could not model an accelerated expansion.
Therefore we may suggest the decay of the EmU with dark energy
domination during the recent history of the Universe. It will be
possible to use the approach suggested in \cite{arrche12gc}, viz.,
to consider a chiral cosmological model coupled to a
perfect fluid with the aim of considering cold dark matter.

We would like to stress once again that with the exact solutions
of the EmU, there is the possibility of calculating the cosmological
parameters  \cite{chefom08} without attracting a slow
roll approximation which is often used for this purpose. We presented
the results of a such a calculation and in future, we hope to obtain
limitations for the EmU model parameters from observational data.

\section{Acknowledgments}

SVC is thankful to the University of KwaZulu-Natal, the University
of Zululand and  the NRF for financial support and warm
hospitality during his visit in 2012 to South Africa. SDM
acknowledges that this work is based upon research supported by
the South African Research Chair Initiative of the Department of
Science and Technology and the National Research Foundation.
\end{document}